\documentclass[prd,preprint]{revtex4}
\usepackage{amssymb}
\usepackage{epsfig}

\begin{document}

\title{Supersymmetric Composite Models on Intersecting D-branes}
\author{Noriaki Kitazawa}
\email{kitazawa@phys.metro-u.ac.jp}
\affiliation{
Department of Physics, Tokyo Metropolitan University,
Hachioji, Tokyo 192-0397, Japan}
\date{\today}

\begin{abstract}
We construct supersymmetric composite models of quarks and leptons
 from type IIA ${\bf T^6}/({\bf Z_2} \times {\bf Z_2})$ orientifolds
 with intersecting D6-branes.
In case of
 ${\bf T^6} = {\bf T^2} \times {\bf T^2} \times {\bf T^2}$
 with no tilted ${\bf T^2}$,
 a composite model of supersymmetric SU$(5)$ grand unified theory
 with three generations is constructed.
In case of that one ${\bf T^2}$ is tilted,
 a composite model with SU$(3)_c \times$SU$(2)_L \times$U$(1)_Y$
 gauge symmetry with three generations is constructed.
These models are not realistic,
 but contain fewer additional exotic particles
 and U$(1)$ gauge symmetries
 due to the introduction of the compositeness
 of quarks and leptons.
The masses of some exotic particles are naturally generated
 through the Yukawa interactions among ``preons''.
\end{abstract}

\preprint{}

\vspace*{3cm}

\maketitle

\section{Introduction}
\label{sec:intro}

The particle models
 on the intersecting D-branes in superstring theories
 (see Refs.\cite{BGKL,AFIRU,CSU} for the essential idea)
 have some advantages
 in comparison with the models based on the field theory.
In addition to the fact that the gravity is naturally included,
 non-gauge interactions
 (Yukawa interactions, higher-dimensional interactions, and so on)
 are naturally introduced in the calculable way, in principle.
Namely, all the interactions are well-determined, in principle.
Therefore, 
 there is a possibility
 to explain the Yukawa coupling hierarchy in the standard model,
 and to justify the assumptions on the higher-dimensional interactions
 in many particle models beyond the standard model.
Especially,
 the models with low-energy supersymmetry
 \cite{CSU,BGO,Honecker,Larosa-Pradisi,CPS,Cvetic-Papadimitriou}
 is interesting,
 because the model can be constructed as a stable solution
 of the superstring theory.
The low-energy supersymmetry can be spontaneously broken
 by the non-perturbative dynamics of gauge interactions\cite{CLW}.
In this paper, we concentrate on supersymmetric models.

The construction of the realistic model is not easy.
The model usually contains
 many additional exotic particles and additional gauge symmetries.
Technically speaking,
 this is due to that many additional D6-branes are required
 in addition to the D6-brane for
 SU$(5)$ or SU$(3)_c \times$SU$(2)_L \times$U$(1)_Y$ gauge symmetry
 to satisfy the tadpole cancellation conditions.
Each additional D6-brane gives additional gauge symmetry.
It is natural to consider that
 the additional gauge interactions may also act some important roles
 in nature.
The composite model of quarks and leptons is one way to be investigated,
 since the confining force requires additional gauge interactions.
In this paper,
 we construct two supersymmetric composite models
 following the prescription given in Ref.\cite{CSU}.

Consider the type IIA superstring theory compactified on
 ${\bf T}^6/({\bf Z}_2 \times {\bf Z}_2)$ orientifold, where
 ${\bf T^6}={\bf T^2} \times {\bf T^2} \times {\bf T^2}$.
The type IIA theory is invariant
 under the ${\bf Z_2} \times {\bf Z_2}$ transformation
\begin{eqnarray}
 \theta: && \qquad X_{\pm}^k \rightarrow e^{\pm i 2\pi v_k} X_{\pm}^k,
\\
 \omega: && \qquad X_{\pm}^k \rightarrow e^{\pm i 2\pi w_k} X_{\pm}^k,
\end{eqnarray}
 where $v=(0,0,1/2,-1/2,0)$ and $w=(0,0,0,1/2,-1/2)$ and
\begin{equation}
 X_{\pm}^k = 
  \left\{
   \begin{array}{ll}
    {1 \over \sqrt{2}} \left( \pm X^{2k} + X^{2k+1} \right),
     & \quad \mbox{for $k=0$}, \\
    {1 \over \sqrt{2}} \left( X^{2k} \pm i X^{2k+1} \right),
     & \quad \mbox{for $k=1,2,3,4$}
   \end{array}
  \right.
\end{equation}
 with space-time coordinates $X^\mu$, $\mu=0,1,\cdots,9$.
The type IIA theory
 is also invariant under the $\Omega R$ transformation,
 where $\Omega$ is the world-sheet parity transformation and
\begin{equation}
 R: \qquad\quad
  \left\{
   \begin{array}{ll}
    X^i \rightarrow X^i,
     & \quad \mbox{for $i=0,1,2,3,4,6,8$}, \\
    X^j \rightarrow -X^j,
     & \quad \mbox{for $j=5,7,9$}.
   \end{array}
  \right.
\end{equation}
We mod out this theory
 by the action of $\theta$, $\omega$, $\Omega R$
 and their independent combinations.

A D6${}_a$-brane stretching over our three-dimensional space
 and winding in compact
 ${\bf T^2} \times {\bf T^2} \times {\bf T^2}$ space
 is specified by the winding numbers in each torus:
\begin{equation}
 [(n_a^1, m_a^1), (n_a^2, m_a^2), (n_a^3, m_a^3)].
\end{equation}
A D6${}_a$-brane is always accompanied by its orientifold image
 D6${}_{a'}$ whose winding numbers are
\begin{equation}
 [(n_a^1, -m_a^1), (n_a^2, -m_a^2), (n_a^3, -m_a^3)].
\end{equation}
The number of intersection
 between D6${}_a$-brane and D6${}_b$-brane is given by
\begin{equation}
 I_{ab} = \prod_{i=1}^3 \left( n_a^i m_b^i - m_a^i n_b^i \right).
\end{equation}
The intersecting angles, $\theta_a^i$,
 in each torus between D6${}_a$-brane and $X^4$, $X^6$ and $X^8$ axes
 are described as
\begin{equation}
 \theta_a^i = \tan^{-1} \left( \chi_i {{m_a^i} \over {n_a^i}} \right),
\end{equation}
 where $\chi_i$ are the ratios of two radii of each torus
 (complex structure moduli).
The system is supersymmetric,
 if $\theta_a^1+\theta_a^2+\theta_a^3=0$ is satisfied for all $a$.
The configuration of intersecting D6-branes should satisfy
 the following tadpole cancellation conditions.
\begin{eqnarray}
 \sum_a N_a n_a^1 n_a^2 n_a^3 &=& 16,
\\
 \sum_a N_a n_a^1 m_a^2 m_a^3 &=& -16,
\\
 \sum_a N_a m_a^1 n_a^2 m_a^3 &=& -16,
\\
 \sum_a N_a m_a^1 m_a^2 n_a^3 &=& -16,
\end{eqnarray}
 where $N_a$ is the multiplicity of D6${}_a$-brane,
 and we are assuming three rectangular (untilted) tori.
In case of some tori are tilted,
 the tadpole cancellation conditions are modified.
For example, in case of the third torus is tilted,
 the conditions become
\begin{eqnarray}
 \sum_a N_a n_a^1 n_a^2 n_a^3 &=& 16,
\\
 \sum_a N_a n_a^1 m_a^2 {\tilde m}_a^3 &=& -8,
\\
 \sum_a N_a m_a^1 n_a^2 {\tilde m}_a^3 &=& -8,
\\
 \sum_a N_a m_a^1 m_a^2 n_a^3 &=& -16,
\end{eqnarray}
 where ${\tilde m}_a^3 \equiv m_a^3 + n_a^3/2$.

There are four sectors of open string corresponding to
 on which D6-branes two ends of open string are fixed:
 $aa$, $ab+ba$, $ab'+b'a$, $aa'+a'a$ sectors.
Each sector gives the matter fields
 in low-energy four-dimensional space-time.
The general massless field contents
 are given in Table \ref{general-spectrum}.
\begin{table}
 \begin{tabular}{|c|l|}
  \hline
  sector & field \\
  \hline\hline
  $aa$   & U$(N_a/2)$ or USp$(N_a)$ gauge multiplet. \\
         & 3 U$(N_a/2)$ adjoint or 3 USp$(N_a)$ anti-symmetric tensor
           chiral multiplets. \\
  \hline
  $ab+ba$ & $I_{ab}$ $( \Box_a, {\bar \Box}_b )$ chiral multiplets. \\
  \hline
  $ab'+b'a$ & $I_{ab'}$ $( \Box_a, \Box_b )$ chiral multiplets. \\
  \hline
  $aa'+a'a$ & ${1 \over 2}
                \left( I_{aa'} - {4 \over {2^k}} I_{aO6} \right)$
                symmetric tensor chiral multiplets. \\
            & ${1 \over 2}
                \left( I_{aa'} + {4 \over {2^k}} I_{aO6} \right)$
                anti-symmetric tensor chiral multiplets. \\
  \hline
 \end{tabular}
\caption{
General massless field contents on intersecting D6-branes.
In $aa$ sector,
 the gauge symmetry is USp$(N_a)$ or U$(N_a/2)$ corresponding to
 whether D6${}_a$-brane is parallel or not
 to some O6-plane, respectively.
In $aa'+a'a$ sector, $k$ is the number of tilted torus,
 and $I_{aO6}$ is the sum of the intersection numbers
 between D6${}_a$-brane and all O6-planes.
}
\label{general-spectrum}
\end{table} 

In the next section
 we construct a composite supersymmetric SU$(5)$ grand unified model
 of Ref.\cite{Strassler} on intersecting D-branes.
In section \ref{sec:SSM}
 we construct a composite supersymmetric
 SU$(3)_c \times$SU$(2)_L \times$U$(1)_Y$ model
 of Refs.\cite{Nelson-Strassler,Kitazawa-Okada}.
These two models are not realistic in many points.
Especially,
 they contain additional light exotic particles
 and additional U$(1)$ gauge symmetries.
We will see, however, that
 these models contain relatively fewer number of exotic particles
 and additional U$(1)$ gauge symmetries
 by virtue of the compositeness of quarks and leptons.
We will also see that
 the configuration of intersecting D-branes
 is very simple in each model.
In section \ref{sec:conclusions}
 we present our conclusions.

\section{Composite SU$(5)$ Grand Unified Model}
\label{sec:SGUT}

In the supersymmetric
 composite SU$(5)$ grand unified model of Ref.\cite{Strassler},
 the fields of $10$ in SU$(5)$ are composite
 and the fields of $5$ and $5^*$ in SU$(5)$ are elementary.
The confining forces
 are strong SU$(2)$ interactions for each generation.
The particle contents of one generation are as follows.
\begin{eqnarray}
 \begin{array}{ccc}
         & \mbox{SU}(2) & \mbox{SU}(5) \\
  P      & {\bf 2}      & {\bf 5}      \\
  N      & {\bf 2}      & {\bf 1}      \\
  \Phi_1 & {\bf 1}      & {\bf 5^*}    \\
  \Phi_2 & {\bf 1}      & {\bf 5^*}    \\
 \end{array}
\nonumber
\end{eqnarray}
The fields $P$ and $N$ are ``preons''.
The $PP$ bound state becomes field of $10$ in SU$(5)$,
 and $PN$ bound state becomes field of $5$ in SU$(5)$.
The fact that
 all the fields belong to the fundamental or anti-fundamental
 representation of the gauge group
 is a good feature to realize this model on intersecting D-branes.

\begin{figure}
\begin{center}
\epsfig{file=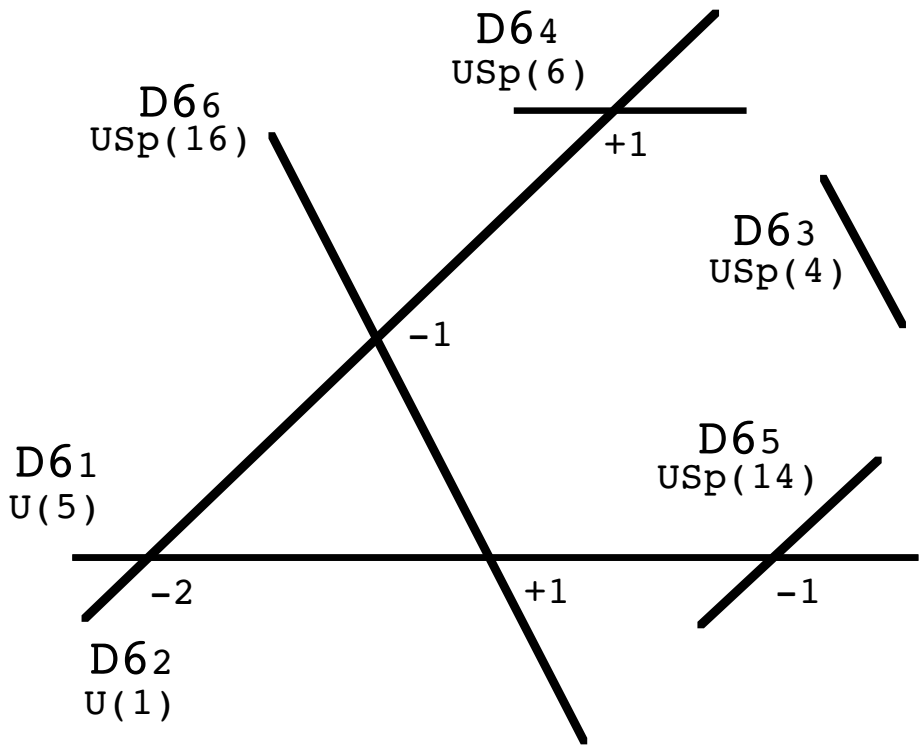,height=100mm}
\end{center}
\caption{
Schematic picture of
 the configuration of the intersecting D6-branes
 for composite supersymmetric SU$(5)$ grand unified model.
This picture describes
 only the situation of the intersection of D6-branes each other,
 and the relative place of each D6-brane has no meaning.
The number at the intersection point
 between D6${}_a$ and D6${}_b$ branes
 denotes intersection number $I_{ab}$ with $a<b$.
}
\label{picture-SGUT}
\end{figure}

The configuration of the intersecting D6-branes
 for this composite model is given in Table \ref{D-brane-SGUT}.
\begin{table}
 \begin{tabular}{|c|c|c|}
  \hline
  D6-brane & winding number & multiplicity     \\
  \hline\hline
  D6${}_1$   & $[(1,0), (1,-1), (1,1)]$ & $10$ \\
  \hline
  D6${}_2$   & $[(1,1), (1,0), (1,-1)]$ & $2$  \\
  \hline
  D6${}_3$   & $[(1,0), (1,0), (1,0)]$  & $4$  \\
  \hline
  D6${}_4$   & $[(1,0), (0,1), (0,-1)]$ & $6$  \\
  \hline
  D6${}_5$   & $[(0,1), (1,0), (0,-1)]$ & $14$ \\
  \hline
  D6${}_6$   & $[(0,1), (0,-1), (1,0)]$ & $16$ \\
  \hline
 \end{tabular}
\caption{
Configuration of the intersecting D6-branes
 for supersymmetric composite SU$(5)$ grand unified model.
All three tori are rectangular (untilted).
Four D6-branes, D6${}_3$, D6${}_4$, D6${}_5$ and D6${}_6$,
 are on top of some O6-planes.
}
\label{D-brane-SGUT}
\end{table}
This configuration is supersymmetric when $\chi_1=\chi_2=\chi_3$,
 and satisfies the tadpole cancellation conditions.
A schematic picture of the configuration
 is given in Fig. \ref{picture-SGUT}.
On the D6${}_1$-brane
 the grand unified gauge group SU$(5)$ is realized.
On the D6${}_2$-brane
 we have U$(1)_{D6_2}$ gauge symmetry
 whose anomaly is cancelled by the Green-Schwarz mechanism,
 and its gauge boson has a mass of the order of the string scale. 
In this configuration
 we have only $aa$ and $ab+ba$ sectors,
 and $ab'+b'a$ and $aa'+a'a$ sectors do not appear.

The gauge symmetries of the D6-branes
 which are on top of some O6-planes can be reduced
 in the following way\cite{CSU,Cvetic-Papadimitriou}.
\begin{eqnarray}
 \mbox{USp}(16) &\longrightarrow&
  \mbox{USp}(2)_1 \times \mbox{USp}(2)_2 \times \mbox{USp}(2)_3
  \times \left( \mbox{U}(1) \right)^3,
\label{usp2_gut}
\\
 \mbox{USp}(14) &\longrightarrow& \left( \mbox{U}(1) \right)^4,
\\
 \mbox{USp}(6) &\longrightarrow& \left( \mbox{U}(1) \right)^2.
\end{eqnarray}
This can be done by moving D6-branes away from O6-planes
 in a way consistent with the orientifold projections.
For the reduction of Eq.(\ref{usp2_gut}),
 we have to assume some non-trivial vacuum expectation values
 of the fields in the anti-symmetric tensor representation
 in $aa$ sector.
Each USp$(2)$ gauge interaction
 corresponds to the additional strong interaction, ``hypercolor'',
 for the confinement of ``preons'' in each generation.
(See Ref.\cite{Intriligator-Pouliot} for the dynamics
 of USp$(N)$ supersymmetric gauge theories.)
The low-energy particle contents before ``hypercolor'' confinement
 is given in Table \ref{contents-SGUT-1}.
\begin{table}
 \begin{tabular}{|c|c|c|}
  \hline
  sector             & $\mbox{SU}(5)
                        \times \mbox{USp}(2)_1
                        \times \mbox{USp}(2)_2
                        \times \mbox{USp}(2)_3$
                                                   & field  \\
  \hline\hline
  $D6_1 \cdot D6_2$  & $(5^*, 1, 1, 1) \times 2$   &
                                           ${\bar H}_{1,2}$ \\
  \hline
  $D6_1 \cdot D6_5$  & $(5^*, 1, 1, 1) \times 14$  &        \\
  \hline
  $D6_1 \cdot D6_6$  & $(5, 2, 1, 1)$              & $P_1$  \\
                     & $(5, 1, 2, 1)$              & $P_2$  \\
                     & $(5, 1, 1, 2)$              & $P_3$  \\
                     & $(5, 1, 1, 1) \times 10$    &        \\
  \hline
  $D6_2 \cdot D6_4$  & $(1, 1, 1, 1) \times 6$     &        \\
  \hline
  $D6_2 \cdot D6_6$  & $(1, 2, 1, 1)$              & $N_1$  \\
                     & $(1, 1, 2, 1)$              & $N_2$  \\
                     & $(1, 1, 1, 2)$              & $N_3$  \\
                     & $(1, 1, 1, 1) \times 10$    &        \\
  \hline
 \end{tabular}
\caption{
Low-energy particle contents
 of composite SU$(5)$ grand unified model
 before ``hypercolor'' confinement.
The fields from $aa$ sectors are neglected for simplicity.
}
\label{contents-SGUT-1}
\end{table}
The fields $P_i$ and $N_i$ with $i=1,2,3$ are ``preons''
 in first, second and third generations, respectively.

The value of the U$(N)$ gauge coupling $g$
 at the string scale is determined by
\begin{equation}
 g^2 \kappa_4^{-1} = \sqrt{4\pi} M_s {\sqrt{V_6} \over {V_3}},
\end{equation}
 where $\kappa_4=\sqrt{8 \pi G_N}$,
 $M_s = 1 / \sqrt{\alpha'}$,
 $V_6$ is the volume of compact six-dimensional space
 and $V_3$ is the volume of corresponding D6-brane
 in compact six-dimensional space\cite{CLW}.
In case of that the D6-brane is parallel to some O6-plane,
 the corresponding USp$(N)$ gauge coupling is given by
\begin{equation}
 g^2 \kappa_4^{-1} = 2 \sqrt{4\pi} M_s {\sqrt{V_6} \over {V_3}},
\end{equation}
 which has an additional factor $2$\cite{BLS}.
In our model,
 the ``hypercolor'' USp$(2)$ has the largest coupling
 at the string scale
 because of the smallest $V_3$ and the additional factor $2$.

As it is easily understood from Fig. \ref{picture-SGUT},
 we have Yukawa couplings among fields in
 D6${}_1 \cdot$D6${}_2$, D6${}_1 \cdot$D6${}_6$
 and D6${}_2 \cdot$D6${}_6$ sectors\cite{CLS}.
There are following Yukawa interactions:
\begin{eqnarray}
 g_1 {\bar H}_1 P_i N_i,
\label{Yukawa-SGUT1}
\\
 g_2 {\bar H}_2 P_i N_i
\label{Yukawa-SGUT2}
\end{eqnarray}
 with $i=1,2,3$.
The Yukawa coupling $g_2$ is exponentially smaller
 than the Yukawa coupling $g_1 \sim 1$.
These interactions become mass terms
 after ``hypercolor'' confinement.

The particle contents after ``hypercolor'' confinement
 is given in Table \ref{contents-SGUT-2}.
\begin{table}
 \begin{tabular}{|c|c|c|}
  \hline
  sector             & $\mbox{SU}(5) \times \mbox{U}(1)_{D6_2}$
                     & field \\
  \hline\hline
  $D6_1 \cdot D6_2$  & $(5^*, 1) \times 2$  
                     & ${\bar H}_{1,2}$ \\
  \hline
  $D6_1 \cdot D6_5$  & $(5^*, 0) \times 14$
                     & $\Phi_{1,2,3}$ and exotics \\
  \hline
  $D6_1 \cdot D6_6$ and $D6_2 \cdot D6_6$
                     & $(10, 0) \times 3$   
                     & $\Sigma_{1,2,3} \sim [PP]_i$ \\
                     & $(5, -1) \times 3$   
                     & $H_{1,2,3} \sim [PN]_i$ \\
                     & $(5, 0) \times 10$   
                     & exotics              \\
                     & $(1, -1) \times 10$  
                     & singlets             \\
  \hline
  $D6_2 \cdot D6_4$  & $(1, 1) \times 6$    
                     & singlets             \\
  \hline
 \end{tabular}
\caption{
Particle contents
 of composite SU$(5)$ grand unified model
 after ``hypercolor'' confinement.
The fields from $aa$ sectors are neglected for simplicity.
}
\label{contents-SGUT-2}
\end{table}
We have three pairs of $10$ and $5^*$
 ($\Sigma$ and $\Phi$, or quarks and leptons) in SU$(5)$,
 one massive pair of $5$ and $5^*$,
 twelve massless pairs of $5$ and $5^*$ (Higgs multiplets),
 and sixteen singlets.
We also have three SU$(5)$ adjoint fields
 and many singlet fields in $aa$ sector.
It is remarkable that
 odd number of generations is realized
 in this simple D6-brane configuration
 with three untilted tori.
There are eleven additional U$(1)$ gauge symmetries
 (two of them are anomalous).

A supersymmetric mass of
 one pair of $5$ and $5^*$ in SU$(5)$
 are generated through the Yukawa couplings
 of Eqs.(\ref{Yukawa-SGUT1}) and (\ref{Yukawa-SGUT2}),
 since the three composite operators $P_i N_i$
 are replaced by three fields $H_i$.
The value of the mass is of the order of
 the scale of dynamics of USp$(2)$.

The Yukawa couplings
 for the mass generation of up-type quarks,
 $\Sigma_i \Sigma_i H_i$ with $i=1,2,3$,
 are dynamically generated.
The values of the coupling constants are of the order of unity,
 which is appropriate to explain large top quark mass.

The D6${}_3 \cdot$D6${}_3$ sector is the hidden sector,
 since D6${}_3$-brane has no intersections with other D6-branes.
Only the interactions
 which are mediated by closed string states
 connect this sector with the other sectors.
If the dynamics of USp$(4)$ gauge interaction
 with three chiral multiplets in anti-symmetric tensor representation
 and with supergravity fields from closed string states
 dynamically breaks supersymmetry,
 the ``sequestering scenario''
 of the mediation of supersymmetry breaking\cite{Randall-Sundrum}
 is naturally realized.

In the next section
 we construct a composite SU$(3)_c \times$SU$(2)_L \times$U$(1)_Y$
 model with fewer exotic particles and U$(1)$ gauge symmetries.

\section{Composite SU$(3)_c \times$SU$(2)_L \times$U$(1)_Y$ Model}
\label{sec:SSM}

\begin{figure}
\begin{center}
\epsfig{file=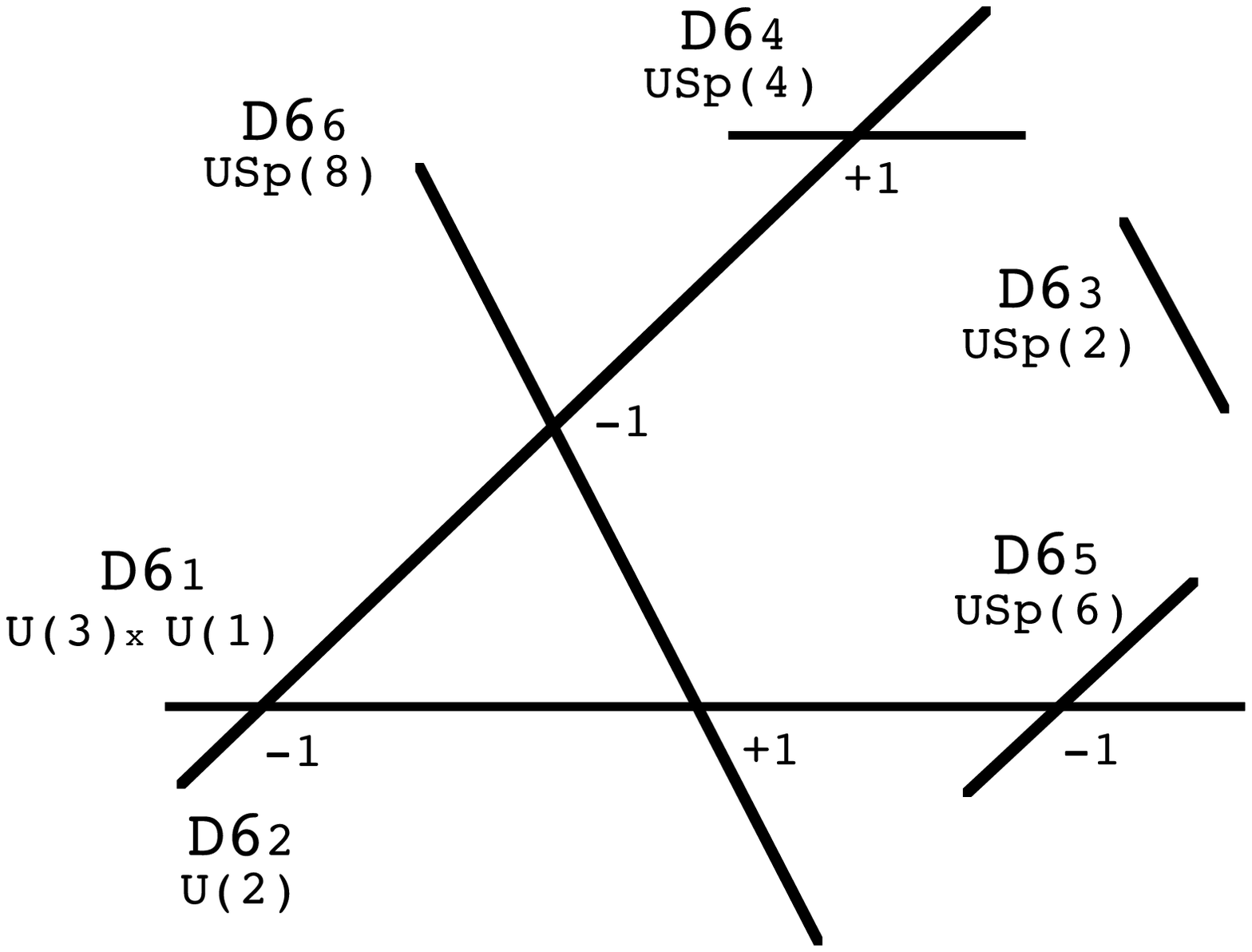,height=100mm}
\end{center}
\caption{
Schematic picture
 of the configuration of the intersection of D6-branes
 for composite supersymmetric
 SU$(3)_c \times$SU$(2)_L \times$U$(1)_Y$ model.
This picture describes
 only the situation of the intersection of D6-branes each other,
 and the relative place of each D6-brane has no meaning.
The number at the intersection point
 between D6${}_a$ and D6${}_b$ branes
 denotes intersection number $I_{ab}$ with $a<b$.
}
\label{picture-SSM}
\end{figure}

The particle contents of one generation of the composite model
 of Refs.\cite{Nelson-Strassler,Kitazawa-Okada} are as follows.
\begin{eqnarray}
 \begin{array}{ccccc}
             & \mbox{SU}(2)_H & \mbox{SU}(3)_c & \mbox{SU}(2)_L &
                                                 \mbox{U}(1)_Y  \\
  C          & {\bf 2}        & {\bf 3}        & {\bf 1}        &
                                                 -{1 \over 3}   \\
  D          & {\bf 2}        & {\bf 1}        & {\bf 2}        &
                                                  {1 \over 2}   \\
  N          & {\bf 2}        & {\bf 1}        & {\bf 1}        &
                                                  0             \\
  {\bar d}   & {\bf 1}        & {\bf 3^*}      & {\bf 1}        &
                                                  {1 \over 3}   \\
  l          & {\bf 1}        & {\bf 1}        & {\bf 2}        &
                                                 -{1 \over 2}   \\
  {\bar \nu} & {\bf 1}        & {\bf 1}        & {\bf 1}        &
                                                  0             \\
  {\bar\Phi} & {\bf 1}        & {\bf 3^*}      & {\bf 1}        &
                                                  {1 \over 3}   \\
  {\bar H}   & {\bf 1}        & {\bf 1}        & {\bf 2}        &
                                                 -{1 \over 2}
 \end{array}
\nonumber
\end{eqnarray}
The fields $C$, $D$ and $N$ are ``preons''.
The bound states of $CC$, $CD$, $DD$, $CN$ and $DN$
 become a right-handed up-type quark, a quark doublet,
 a right-handed charged lepton, a colored Higgs
 and a Higgs doublet, respectively.
We construct a composite model similar to this model
 (the hypercharge assignment will be different)
 on intersecting D6-branes.
We assume that
 only the third torus is tilted.
The intersecting D6-brane configuration
 is given in Table \ref{D-brane-SSM}.
\begin{table}
 \begin{tabular}{|c|c|c|}
  \hline
  D6-brane & winding number & multiplicity        \\
  \hline\hline
  D6${}_1$   & $[(1,0), (1,-1), (1,1/2)]$ & $6+2$ \\
  \hline
  D6${}_2$   & $[(1,1), (1,0), (1,-1/2)]$ & $4$   \\
  \hline
  D6${}_3$   & $[(1,0), (1,0), (2,0)]$  & $2$     \\
  \hline
  D6${}_4$   & $[(1,0), (0,1), (0,-1)]$ & $4$     \\
  \hline
  D6${}_5$   & $[(0,1), (1,0), (0,-1)]$ & $6$     \\
  \hline
  D6${}_6$   & $[(0,1), (0,-1), (2,0)]$ & $8$     \\
  \hline
 \end{tabular}
\caption{
Configuration of the intersecting D6-branes
 for supersymmetric composite
 SU$(3)_c \times$SU$(2)_L \times$U$(1)_Y$ model.
Only the third torus is tilted.
Four D6-branes, D6${}_3$, D6${}_4$, D6${}_5$ and D6${}_6$,
 are parallel to O6-planes.
D6${}_1$-brane consists of two parallel D6-branes
 with multiplicities $6$ and $2$.
}
\label{D-brane-SSM}
\end{table}
This configuration is supersymmetric when
 $\chi_1:\chi_2:\chi_3=1:1:2$,
 and satisfies the tadpole cancellation conditions.
A schematic picture of the intersection of D6-branes
 is given in Fig. \ref{picture-SSM}.
In this configuration
 we have only $aa$ and $ab+ba$ sectors,
 and $ab'+b'a$ and $aa'+a'a$ sectors do not appear.

On the D6${}_1$-brane
 the gauge symmetry of
 SU$(3)_c \times$U$(1)_c \times$U$(1)$ is realized,
 where SU$(3)_c \times$U$(1)_c$
 comes from the D6${}_1$-brane of multiplicity $6$
 and U$(1)$ comes from the D6${}_1$-brane of multiplicity $2$.
The charges of these two U$(1)$ gauge symmetries
 are denoted as $Q_c$ and $Q$.
The U$(1)$ symmetry
 which is generated by $Q_c/3-Q$ is anomaly free.
On the D6${}_2$-brane the gauge symmetry of
 SU$(2)_L \times$U$(1)_L$ is realized.
The anomaly of U$(1)_L$
 is cancelled by the Green-Schwarz mechanism,
 and the corresponding gauge boson has a mass of the string scale.

The gauge symmetries of the D6-branes
 which are on top of some O6-planes are reduced as follows.
\begin{eqnarray}
 \mbox{D6}_4: &\qquad&
  \mbox{USp}(4) \rightarrow \mbox{U}(1)_{D6_4},
\label{usp4}
\\
 \mbox{D6}_5: &\qquad&
  \mbox{USp}(6) \rightarrow \mbox{USp}(4) \times \mbox{USp}(2)
                \rightarrow \mbox{U}(1)_{D6_5,1}
                            \times \mbox{U}(1)_{D6_5,2},
\\
 \mbox{D6}_6: &\qquad&
  \mbox{USp}(8) \rightarrow \mbox{USp}(2)_1 \times
                            \mbox{USp}(2)_2 \times
                            \mbox{USp}(2)_3 \times
                            \mbox{USp}(2)_4.
\label{usp2_sm}
\end{eqnarray}
This can be done by moving D6-branes away from O6-planes
 in a way consistent with the orientifold projections.
For the reduction of Eq.(\ref{usp2_sm}),
 we have to assume some non-trivial vacuum expectation values
 of the fields in the anti-symmetric tensor representation
 in $aa$ sector.
All these three U$(1)$ gauge symmetries are anomaly free.
The charges of these U$(1)$ gauge symmetries are denoted as
 $Q_{D6_4}$, $Q_{D6_5,1}$ and $Q_{D6_5,2}$.
Four USp$(2)$ gauge interactions are
 all strong coupling ``hypercolor'' gauge interactions
 with the largest coupling constants at the string scale.
D6${}_3$-brane with USp$(2)$ gauge symmetry is the hidden sector,
 since it has no intersections with other D6-branes.

The low-energy particle contents before ``hypercolor'' confinement
 is given in Table \ref{contents-SSM-1}.
\begin{table}
 \begin{tabular}{|c|c|c|}
  \hline
  sector             & $\mbox{SU}(3)_c \times \mbox{SU}(2)_L
                       \times \mbox{USp}(2)_1
                       \times \mbox{USp}(2)_2
                       \times \mbox{USp}(2)_3
                       \times \mbox{USp}(2)_4$
                     & field  \\
                     & ($Q_c/3-Q, Q_{D6_4}, Q_{D6_5,1}, Q_{D6_5,2}$)
                     &        \\
  \hline\hline
  $D6_1 \cdot D6_2$  & $(3^*, 2, 1, 1, 1, 1)_{(-1/3,0,0,0)}$
                     & ${\bar q}$ \\
                     & $(1, 2, 1, 1, 1, 1)_{(+1,0,0,0)}$
                     & ${\bar l}$    \\
  \hline
  $D6_1 \cdot D6_5$  & $(3^*, 1, 1, 1, 1, 1)_{(-1/3,0,\pm1,0)}
                        \times 2$
                     &        \\
                     & $(3^*, 1, 1, 1, 1, 1)_{(-1/3,0,0,\pm1)}$
                     &        \\
                     & $(1, 1, 1, 1, 1, 1)_{(+1,0,\pm1,0)}
                        \times 2$
                     &        \\
                     & $(1, 1, 1, 1, 1, 1)_{(+1,0,0,\pm1)}$
                     &        \\
  \hline
  $D6_1 \cdot D6_6$  & $(3, 1, 2, 1, 1, 1)_{(1/3,0,0,0)}$
                     & $C_1$  \\
                     & $(3, 1, 1, 2, 1, 1)_{(1/3,0,0,0)}$
                     & $C_2$  \\
                     & $(3, 1, 1, 1, 2, 1)_{(1/3,0,0,0)}$
                     & $C_3$  \\
                     & $(3, 1, 1, 1, 1, 2)_{(1/3,0,0,0)}$
                     & $C_4$  \\
                     & $(1, 1, 2, 1, 1, 1)_{(-1,0,0,0)}$
                     & $N_1$  \\
                     & $(1, 1, 1, 2, 1, 1)_{(-1,0,0,0)}$
                     & $N_2$  \\
                     & $(1, 1, 1, 1, 2, 1)_{(-1,0,0,0)}$
                     & $N_3$  \\
                     & $(1, 1, 1, 1, 1, 2)_{(-1,0,0,0)}$
                     & $N_4$  \\
  \hline
  $D6_2 \cdot D6_4$  & $(1, 2, 1, 1, 1, 1)_{(0,\pm1,0,0)}
                        \times 2$
                     &        \\
  \hline
  $D6_2 \cdot D6_6$  & $(1, 2, 2, 1, 1, 1)_{(0,0,0,0)}$
                     & $D_1$  \\
                     & $(1, 2, 1, 2, 1, 1)_{(0,0,0,0)}$
                     & $D_2$  \\
                     & $(1, 2, 1, 1, 2, 1)_{(0,0,0,0)}$
                     & $D_3$  \\
                     & $(1, 2, 1, 1, 1, 2)_{(0,0,0,0)}$
                     & $D_4$  \\
  \hline
 \end{tabular}
\caption{
Low-energy particle contents of
 SU$(3)_c \times$SU$(2)_L \times$U$(1)_Y$ model
 before ``hypercolor'' confinement.
The fields from $aa$ sectors are neglected for simplicity.
}
\label{contents-SSM-1}
\end{table}
The fields $C_i$, $D_i$ and $N_i$ with $i=1,2,3,4$ are ``preons''
 in first, second, third and fourth generations, respectively.
It is easy to understand that
 there are Yukawa interactions of
\begin{eqnarray}
 &{\bar q} C_i D_i,&
\label{Yukawa-1}
\\
 &H N_i D_i&
\label{Yukawa-2}
\end{eqnarray}
 with $i=1,2,3,4$.
All the coupling constants of these Yukawa interactions
 are of the order of unity.
These Yukawa interactions become mass terms of the exotics
 ${\bar q}$ and ${\bar l}$ after ``hypercolor'' confinement.

The particle contents after ``hypercolor'' confinement
 is given in Table \ref{contents-SSM-2}.
\begin{table}
 \begin{tabular}{|c|c|c|}
  \hline
  sector             & $\mbox{SU}(3)_c \times \mbox{SU}(2)_L
                        \times$U$(1)_Y$
                     & field \\
  \hline\hline
  $D6_1 \cdot D6_2$  & $(3^*, 2)_{-1/6}$  
                     & ${\bar q}$ \\
                     & $(1, 2)_{1/2}$  
                     & ${\bar l}$ \\
  \hline
  $D6_1 \cdot D6_5$  & $(3^*, 1)_{1/3} \times 3$
                     & ${\bar d}_{1,2,3}$ \\
                     & $(3^*, 1)_{-2/3} \times 3$
                     & ${\bar u}_{1,2,3}$ \\
                     & $(1, 1)_{0} \times 3$
                     & ${\bar \nu}_{1,2,3}$ \\
                     & $(1, 1)_{1} \times 3$
                     & ${\bar e}_{1,2,3}$ \\
  \hline
  $D6_1 \cdot D6_6$ and $D6_2 \cdot D6_6$
                     & $(3^*, 1)_{1/3} \times 4$   
                     & ${\bar d}'_{1,2,3,4} \sim [CC]_i$ \\
                     & $(3, 2)_{1/6} \times 4$   
                     & $q_{1,2,3,4} \sim [CD]_i$ \\
                     & $(1, 1)_{0} \times 4$   
                     & $S_{1,2,3,4} \sim [DD]_i$ \\
                     & $(3, 1)_{-1/3} \times 4$   
                     & $\Phi_{1,2,3,4} \sim [CN]_i$ \\
                     & $(1, 2)_{-1/2} \times 4$   
                     & $l_{1,2,3,4} \sim [DN]_i$ \\
  \hline
  $D6_2 \cdot D6_4$  & $(1, 2)_{1/2} \times 2$    
                     & $H_{1,2}$ \\
                     & $(1, 2)_{-1/2} \times 2$    
                     & ${\bar H}_{1,2}$ \\
  \hline
 \end{tabular}
\caption{
The particle contents of composite
 SU$(3)_c \times$SU$(2)_L \times$U$(1)_Y$ model
 after ``hypercolor'' confinement.
The fields from $aa$ sectors are neglected for simplicity.
}
\label{contents-SSM-2}
\end{table}
The hypercharge is defined as
\begin{equation}
 {Y \over 2} = {1 \over 2} \left( {{Q_c} \over 3} - Q \right)
             + {1 \over 2} \left(
                            Q_{D6_4} +
                            Q_{D6_5,1} + Q_{D6_5,2}
                           \right).
\end{equation}
We can see that
 the particle contents of three generations of
 quarks and leptons are included.
The hypercharge assignment of ``preons''
 are different from that in the model of
 Refs.\cite{Nelson-Strassler,Kitazawa-Okada}.

The exotic ${\bar q}$ gets a mass
 with one linear combination of $q_{1,2,3,4}$
 through the Yukawa coupling of Eq.(\ref{Yukawa-1}),
 and three left-handed doublet quarks remain massless.
The exotic ${\bar l}$ gets a mass
 with one linear combination of $l_{1,2,3,4}$
 through the Yukawa coupling of Eq.(\ref{Yukawa-2}),
 and three left-handed doublet leptons remain massless.
The values of these masses are of the order of
 the scale of dynamics of USp$(2)$.

The following Yukawa interactions are dynamically generated
 with the coupling constants of the order of unity.
\begin{equation}
 {\bar d}'_i q_i l_i,
\qquad
 {\bar d}'_i \Phi_i S_i,
\qquad
 q_i q_i \Phi_i
\end{equation}
 with $i=1,2,3,4$.
If singlets $S_i$ have vacuum expectation value,
 the exotics of ${\bar d}'_i$ and $\Phi_i$ become massive.
There is no mechanism for giving mass
 to two pairs of Higgs doublets, $H_{1,2}$ and ${\bar H}_{1,2}$.

Three SU$(3)_c$ adjoint chiral multiplets,
 three SU$(2)_L$ adjoint chiral multiplets,
 and many singlets in $aa$ sector remain massless.
In addition to the standard model gauge symmetry,
 there are five additional U$(1)$ gauge symmetries
 (two of them are anomalous).

The D6${}_3 \cdot$D6${}_3$ sector is the hidden sector,
 since D6${}_3$-brane has no intersections with other D6-branes.
Only the interactions
 which are mediated by closed string states
 connect this sector with the other sectors.
The system of this sector is
 the supersymmetric USp$(2)$ Yang-Mills theory
 with supergravity fields from closed string states.
If the supersymmetry is dynamically broken in this sector,
 the ``sequestering scenario''
 of the mediation of supersymmetry breaking is naturally realized.

This model has almost the same features
 of the model constructed in the previous section,
 but contain fewer number of
 the exotic particles and additional U$(1)$ gauge symmetries.

\section{Conclusions}
\label{sec:conclusions}

We have constructed two composite models of quarks and leptons
 from type IIA ${\bf T^6}/({\bf Z_2} \times {\bf Z_2})$ orientifolds
 with intersecting D6-branes.
The configuration of the intersecting D6-branes of each model
 is very simple.
The number of
 the exotic particles and additional U$(1)$ gauge symmetries
 is relatively fewer than in the models previously constructed
 by virtue of the compositeness of quarks and leptons.
Since the compositeness requires
 additional physical gauge interactions, ``hypercolors'',
 the number of additional unphysical gauge symmetries
 is reduced.
Once the number of
 additional unphysical gauge symmetries is reduced,
 the number of additional exotic particles is also reduced,
 since the exotic particles are required for
 the anomaly cancellation of the additional gauge symmetries.

One interesting feature in these models
 is the generation of masses of exotic particles
 through the Yukawa couplings among ``preons''.
This mechanism can be applied
 to give masses to exotic particles in more realistic models.

It is also interesting
 that the hidden or ``sequestering'' sector is naturally emerged
 in these models.
The dynamics in this hidden sector
 might give supersymmetry breaking and a source of the dark energy.

It would be very interesting
 to explorer more realistic models
 from more general intersecting D-brane configurations.

\acknowledgments

This work was supported in part by Yamada Science Foundation
(No. 2003-4022).

\end{document}